\documentclass[%
 reprint,
 amsmath,amssymb,
 aps,
 prd,
floatfix,
showkeys
]{revtex4-1}

\usepackage[USenglish]{babel}
\usepackage{slashed}
\usepackage{graphicx}
\usepackage{dcolumn}
\usepackage{bm}
\usepackage{amssymb}
\usepackage{amsmath}
\usepackage{graphicx}
\usepackage{epsfig}
\usepackage{rotating}

\usepackage[utf8]{inputenc} 
\usepackage{amsmath} 
\usepackage{array}
\usepackage{float}
\usepackage{color}
\usepackage[usenames,dvipsnames]{xcolor}
\usepackage{epstopdf}
\usepackage{natbib}
\usepackage{graphicx}
\usepackage{lineno}
\usepackage{slashed}
\usepackage{color}
\usepackage{hyperref}
\usepackage{float}
\usepackage{mathtools,slashed}
\usepackage{multirow}

\newcommand{\be}{\begin{equation}}
\newcommand{\ee}{\end{equation}}
\newcommand{\bea}{\begin{eqnarray}}
\newcommand{\eea}{\end{eqnarray}}
\newcommand{\beas}{\begin{eqnarray*}}
\newcommand{\eeas}{\end{eqnarray*}}

\newcommand{\bse}{\begin{subequations}}
\newcommand{\ese}{\end{subequations}}

\begin{document}

\title{Characterization in Geant4 of different PET configurations}%segundo parentesis es el nombre

\author{M. L. L\'opez~Toxqui$^a$}
\author{C. H.~Zepeda~Fern\'andez$^{a,b}$}
\email{Corresponding author, email: hzepeda@fcfm.buap.mx}
\author{L. F. Rebolledo Herrera$^a$}
\author{B. De Celis~Alonso$^a$}
\author{E.~Moreno~Barbosa$^a$}

\address{
$^a$Facultad de Ciencias F\'isico Matem\'aticas, Benem\'erita Universidad Aut\'onoma de Puebla, Av. San Claudio y 18 Sur, Ciudad Universitaria 72570, Puebla, Mexico\\
$^b$Cátedra CONACyT, 03940, CdMx Mexico.
%$^3$Hospital de la Mujer Puebla, Antiguo Camino Guadalupe Hidalgo 11350, Agua santa, Guadalupe Hidalgo, 72490 Puebla, Pue.
}

\renewcommand{\thempfootnote}{\arabic{mpfootnote}}
\begin{abstract}
Positron Emission Tomography (PET) is a Nuclear Medicine technique that creates images that allow the study of metabolic activity and organ function using radiopharmaceuticals. Continuous improvement of scintillation detectors for radiation in PET as well as improvement in electronic detectors (e.g., SiPMs) and signal processing, makes the field of PET a fast and changing environment. If industry desires to build new systems implementing these technological improvements, it is of its interest to develop modelling strategies that can provide information on how to build them, reducing time and material costs. Bearing this in mind three different PET configurations, were simulated in Geant4, to determine which one presented the best performance according to quality parameters such as spatial resolution (SR), coincident time resolution (CTR) and acceptance value (A). This was done with three different (in size) pairs of LYSO crystals + SiPM detectors. It was found that the 2 Modules system presented worst results than the two Ring detector configurations. Between the Ring configurations the first was marginally better than the second.
\end{abstract}

\keywords{Positron emission tomography, intrinsic time resolution, coincidence time resolution, spatial resolution, Geant4, simulation.}

\renewcommand{\thesection}{\arabic{section}}
\renewcommand{\thesubsection}{\arabic{section}.\arabic{subsection}}
\maketitle
%\begin{center}
%\textcolor{blue}{Highlights}
%\end{center}
%\begin{itemize}
%\item Obtaining clear images of organs to be analyzed is of utmost importance for diagnosis.
%\item Through a PET scan, the functioning or activity of an organ is known.
%\item The smaller the spatial resolution value, the better the image of the organ will be obtained.
%\item The spatial resolution and the coincidence timing resolution values depend of the radionuclide used.
%\item The geometry of a PET is very important, the more crystals there are, the greater the detection of gammas.
%\end{itemize}

\section{Introduction}
Nuclear Medicine is a specialty of diagnostic medicine used to assess organ function as well as diagnosing the severity of different illnesses. It accomplishes this using radiopharmaceutical (RPs). RPs are built attaching a biological compatible molecule to a radionuclide (RN) which emits different particles or rays that are then detected and measured. RPs are adapted for each body part and illness that needs to be targeted and diagnosed ~\cite{radio,radioinfo,RP3,RP2}. For PET applications (a technique of nuclear medicine), the RNs used to build RPs are based on $\beta^+$ decays: A proton from the atom nucleus decomposes into a neutron plus a positron and an electron neutrino ($p \rightarrow n + e^+ + \nu^-$) and the $e^+$ and $\nu^-$ are then ejected from the RN. The $\nu^-$ does not interact with the matter, i.e., all the atoms of the surrounding organs and therefore can be ignored. However, when the $e^+$ slows down due to electrostatic interaction with the surrounding atoms, it will then interact with a random electron, producing the phenomenon of annihilation in which their mass is converted in energy emissions. Each of the two gamma rays produced then, have an energy of 511 keV and are ejected at $180^o$ from each other. The detection of these two gamma rays allows clinicians to know the place where they originated, and is the basis used by PET to obtain an image ~\cite{pet,pet2,pet3}.\\

The most common detection pair used in PET for gamma ray detection is: Lutetium–yttrium oxyorthosilicate (LYSO) scintillating crystals ~\cite{lyso,lyso2,B10} coupled to Photo-Multiplier Tubes (PMTs). Preclinical PET scanners (or scanners for small species), have smaller bore sizes compared to their human counterparts ~\cite{B11,petanimal3,petanimal2}. Therefore, the crystals must be smaller (order of millimeters) to maintain a proportional spatial resolution to those of human use. Silicon Photomultipliers (SiPM) are different types of photosensors, smaller than PMTs (order of millimeters) ~\cite{BeBe}. For this reason, the use of SiPM instead of PMT in PET is recently being seriously considered ~\cite{B15,B16}, especially for the preclinical system cases. In fact, the use of SiPMs may allow the use of configurations of a single crystal coupled to one SiPM improving spatial resolution, efficiency, and photon time detection ~\cite{B17}.\\

As the electronic capabilities advance, so does the interest in optimizing PET systems (human and preclinical). To this end, new scintillation + detector pairs for PET or micro-PET are under constant development ~\cite{B15,B17,B18}. The process of new developments and designs must always be preceded by simulations that give a theoretical quantification of the performance of the new systems. Geant4 is a software developed to study through Monte Carlo statistics the interaction of radiation with matter \cite{geant4}. It was especially developed for radiotherapy scenarios but can model the interaction of gamma rays with radiation detectors \cite{B20}. Geant4 has been used in the past to model PET hardware and configurations as can be seen in \cite{B21} and \cite{B22}.\\

Finally, to perform these simulations as well as quantifications of new PET systems, a series of established parameters are used to compare the performance of the different PET systems  \cite{B23}:
 
\begin{description}
  \item[\textit{Coincident Time Resolution (CTR)}]
    Time difference between the first crystal that detects one gamma and the diametrically opposite crystal that detects the other gamma due to electron-positron annihilation.
    \item[\textit{Spatial Resolution (SR)}]
    Minimum distance by which two points in a medical image can be differentiated.
    \item[\textit{Acceptance}]
    Defined by the following equation:
    \begin{align}
        \mathcal{A} = \dfrac
        {\text{Number of true events recorded}}{\text{Number of annihilations produced}}
    \end{align}
\end{description}

The work presented here was aimed at simulating new combinations of crystal + SiPM pairs for three different PET configurations using Geant4. This to provide a quantification of which of the three configurations would produce better technical results and therefore, reduce costs and time when building these systems. Each configuration was characterized by measuring the CTR, SR and $\mathcal{A}$ parameters which have been previously introduced, and compared in the case of SR, with different values from the literature \cite{bib9}.

\section{Methodology}

\subsection{Software and Hardware}
Geant4 v.10.7 software \cite{bib4} was used for all simulations. Modelling was run on a commercial laptop ASUS VivoBook model X512FB-BR332T with the following characteristics: Intel\textregistered ~ Core\texttrademark ~ i5-8265U processor, with a frequency of 1.6 GHz, with 4 cores, 12 GB RAM. A total storage capacity of 1128 GB, with an HDD of 1000 GB and a SSD of 120 GB.

\subsection{Pair modellation (crystal + SiPM)}
\begin{figure}[h!]
\centering
\includegraphics[width=0.9\columnwidth]{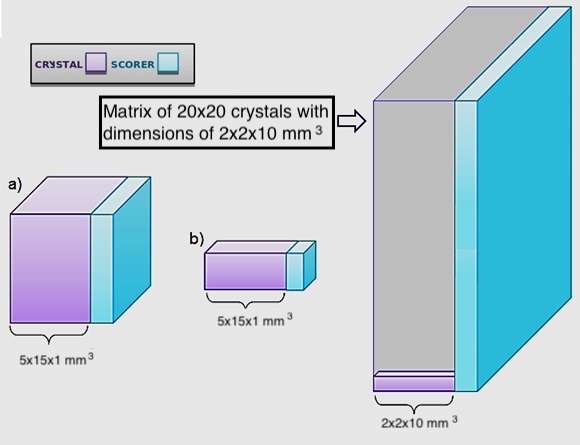}
\caption{Representation of the three different sizes of LYSO crystals coupled to a scorer. LYSO crystal of dimensions were (a) $15 \times 15 \times 10 mm^3$ scenario; (b) $5 \times 5 \times 10 mm^3$ scenario. (c) Array of $20 \times 20$ LYSO crystals. Each crystal has dimensions of $2 \times 2 \times 10 mm^3$ coupled to a scorer with an area of $40 \times 40 mm^2$ scenario.} 
\label{sco}
\end{figure}

For the simulations performed in this paper, the medium surrounding the system was made up by air. To model the LYSO crystals, the elemental composition of the material (presented in Table \ref{tab1}) as well as its optical properties (Table \ref{tab2}) were considered by Geant4 software. However, this platform does not simulate electronics and therefore, it was not possible to simulate the electronic components of a SiPM. For this reason, the LYSO crystals were considered to be coupled to the sensitive detecting area named from now on as scorer, with a $100\%$ efficiency. That scorer represents and replaces the whole SiPM in these models. Figure \ref{sco} shows the three different dimensions of simulated LYSO crystals coupled to a scorer of equal size. LYSO crystal dimensions were: (a) $15 \times 15 \times 10 mm^3$ coupled to scorer with area of $15 \times 15 mm^2$; (b) $5 \times 5 \times 10 mm^3$ coupled to a scorer with area of $5 \times 5 mm^2$. (c) Array of $20 \times 20$ LYSO crystals. Each crystal has dimensions of $2 \times 2 \times 10 mm^3$ coupled to a scorer with an area of $40 \times 40 mm^2$. In all cases the crystal matrix covered the entire scorer sensitive area.

%%%%%%%%%%%%%%%%% TABLA
\begin{table}[htb]
\begin{center}
\begin{tabular}{ c c c }
\hline
Element & Symbol & Percentage \\ \hline
Lutetium & $Lu$ & $71.45\%$ \\
Yttrium & $Y$ & $4.03\%$ \\ 
Silicon & $Si$ & $ 6.37\%$\\
Oxygen & $O$ & $18.15\%$  
\\\hline
\end{tabular}
\caption{Atomic percentage composition of a LYSO. Adapted from \cite{bib4}.}
\label{tab1}
\end{center}
\end{table}

%%%%%%%--------------------------- TABLA
\begin{table}[htbp]
%\begin{center}
\resizebox{8.8cm}{!} {
\begin{tabular}{ c c c c c }
\hline
 Absorption & Wavelength & Attenuation & Refractive & Light output \\ 
 length (cm) & (nm) & length (mm) & index &(photons/keV) \\
 \hline
 $20$ & $420$ & $12$ & $1.82$ &$26$   
\\\hline
\end{tabular}
}
\caption{Optical properties of LYSO crystal. Adapted from \cite{bib5}.}
\label{tab2}
%\end{center}
\end{table}

%%%%%%%%%%%%%%%%%%%%%%%%%%%%%%%%%%% FIGURA

\subsection{PET configurations}
\begin{figure}[htbp]
\centering
\includegraphics[width=0.90\columnwidth]{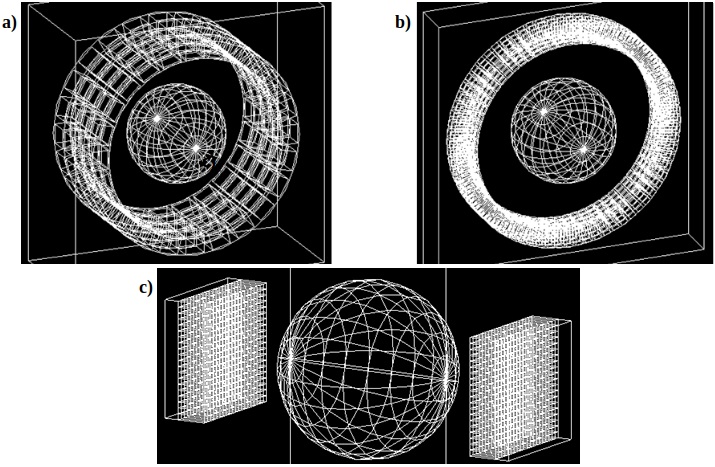}
\caption{
Three PET configurations modelled with Geant4. a) and b) are ring-shaped configurations with LYSO crystals of sizes $15 \times 15 \times 10 mm^3$ and $5 \times 5 \times 10 mm^3$ respectively. The third configuration c) consists of two detector modules composed of a $20 \times 20$ matrix of crystals size $2 \times 2 \times 10 mm^3$. In the center of all simulations, a spheric water phantom in which RNs were dissolved during modellation.} 
\label{petdos}
\end{figure}

The three different PET configurations modelled in this work can be seen in Figure \ref{petdos}, and are described as follows:

\begin{enumerate}
\item Ring-PET: 
Designed by four rings of crystal detectors with a 12 cm diameter. Each ring was made up by 24 LYSO crystals of sizes $15 \times 15 \times 10 mm^3$, giving a grand total of 96 crystals. Separation between crystals in each ring was $75~\mu m$. This PET configuration was based on ~\cite{petanimal3} and is shown in Figure~\ref{petdos} a.

\item Ring-PET$_2$: Designed as the previous configuration with the only difference being the size of the crystals, which in this case were $5 \times 5 \times 10 mm^3$ maintaining the separation of $75~\mu m$ between each other. Each ring was therefore composed of 72 crystals, giving a total of 288 elements. This configuration is presented in Figure~\ref{petdos} b.

\item 2 Module-PET: The design was based on a two module PET instead of the standard ring geometry \cite{bib5}. Each module contained a matrix of $20 \times 20$ LYSO crystals of $2 \times 2 \times 10 mm^3$ with a separation of $75~\mu m$ between them. In total there were 800 crystals. The separation between modules was 10.2 cm. This configuration is presented in Figure~\ref{petdos} c.     
\end{enumerate}

\begin{figure}[H]
\centering
\includegraphics[width=0.9\columnwidth]{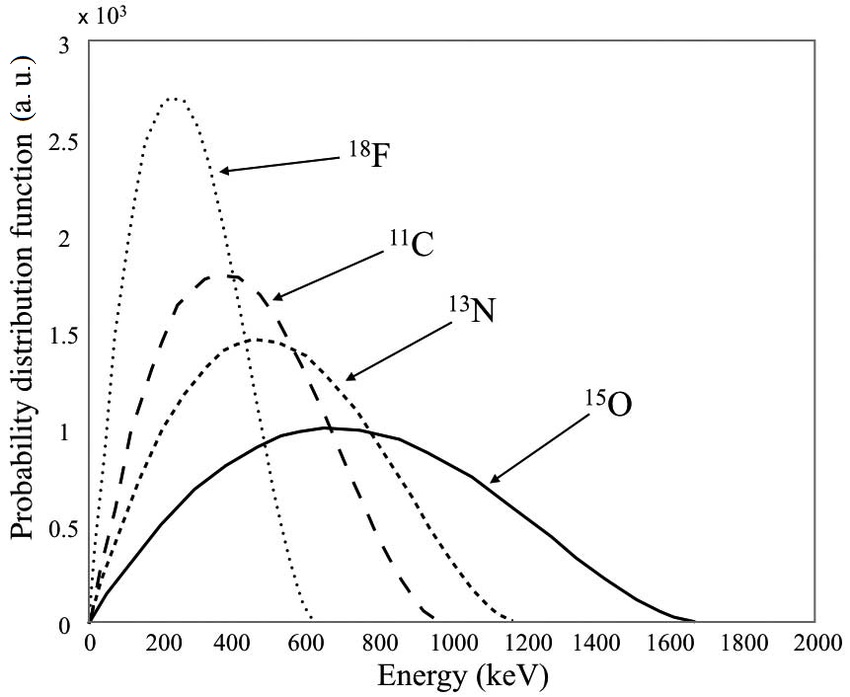}
\caption{Positron kinetic energy spectrum for $^{18}$F, $^{11}$C, $^{13}$N and $^{15}$O. Adapted from ~\cite{bib9}.} 
\label{ff4.3}
\end{figure}

For each of the three PET configurations, 356,000 events for the four RNs ($^{18}$F, $^{11}$C, $^{13}$N and $^{15}$O) were modeled. These are RNs usually employed in PET. The kinetic energy distributions for the different positron sources simulated for this study can be observed in Figure \ref{ff4.3}. A phantom in the shape of a sphere, with a diameter of 6 cm, composed of deionized water was the source of gamma rays in these configurations. The positrons were simulated in each case as a punctual event liberated at the center of the sphere and left to scatter randomly in the radial direction until annihilation. Annihilation would happen when the kinetic energy of each positron was completely lost. Therefore, when positrons were deposited in the center of the sphere, they were deposited with different kinetic energies corresponding to the emitting RN.

\subsection{PET Parameter calculation (CTR, SR and $\mathcal{A}$)}

\begin{figure}[H]
        \centering
        \includegraphics[width=1.\columnwidth]{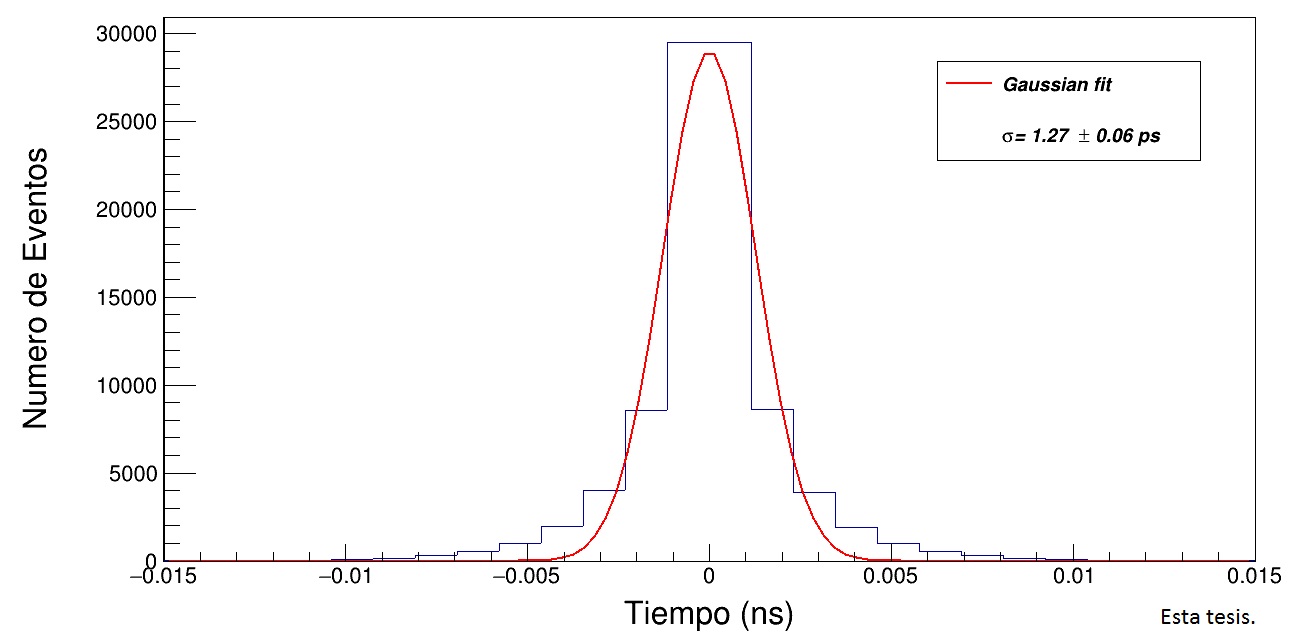}
        \caption{Example of a Photon time-difference distribution used for CTR calculation. Graph obtained from all events detected by the Ring-PET configuration and for $^{18}F$.} 
        \label{vueloF15}
        \end{figure}

To calculate the coincident temporal resolution (CTR), the time difference between photon arrivals, was obtained for all events and for each RNs, and then plotted as an histogram. A Gaussian fit was applied to show the central tendency as seen in Figure \ref{vueloF15}. In which $\sigma$ represents the dispersion of the data round the mean. Sigma has been used in the past to calculate the time resolution of a system \cite{BeBe}.

%%%%%%%%%%%%%%%%%%%%%%%%%%%%%%%%%%%%%%
%%%%%%%%%%%%%%%%%%%%%%%%%%%%%%%%
To calculate the Spatial Resolution (SR), Geant4 provides information on the coordinates (X, Y, Z) of the impact in the crystal of the gamma ray and the flight time (from creation to the first impact). The most important assumption in these calculations was to choose the two gammas that came from the same annihilation. If one gamma was detected in one crystal, the other gamma should be detected in the diametrically opposite crystal. With all this information in mind, it was possible to assume that the angle of their trajectories was approximately $\sim 180 ^o$. Thus, using the geometry and calculating the difference in arrival times between the coincident gammas, it was possible to calculate the coordinates of the annihilation point. To this end, a histogram of the distribution on the X coordinates for the four RNs in the different configurations was plotted. From these distributions, the full width at half maximum (FWHM) and the full width at $1/10$ maximum (FWTM) were obtained. These values determined the minimum separation distance between measured points, that is, the SR. Finally, to obtain the acceptance value of the system ($\mathcal{A}$), the relationship between the number of coincident events detected in each PET configuration by the total number of annihilations produced was calculated. An event consists in the emission of a positron.

\subsection{Statistics}
All tables present in this work show values in the format mean $\pm$  standard deviation. All graphics in figures in this paper present the average value with error bars in the y axis corresponding to standard deviation. When comparing the different parameters of this study, first parametricity was assessed and then once it was confirmed, One way ANOVA tests were performed on data with the three configurations as the ANOVA factor. If ANOVA showed independence between variables, Post-hoc testing (Turkey tests) were calculated after to find differences between pairs of data. A standard $\alpha<$0.05 was required to achieve statistical significance. IBM SPSS statistics version 24 for statistical analysis was used for these calculations.

%\subsubsection{Spatial Resolution}
%\subsubsection{Aceptance}

%----------------
%\subsubsection{Spatial Resolution}

%\begin{figure}[h!]
%\centering
%\includegraphics[width=0.5\columnwidth]{fig4.png}
%\caption{Interaction of positrons in the water sphere. The green lines represent the gammas comming from the annihilation points. The full crystals represent the detection of a gamma particle.} 
%\label{ff4.4}
%\end{figure}

\section{Results}
\subsection{CTR}
\begin{figure}[H]
\centering
\includegraphics[width=1.0\columnwidth]{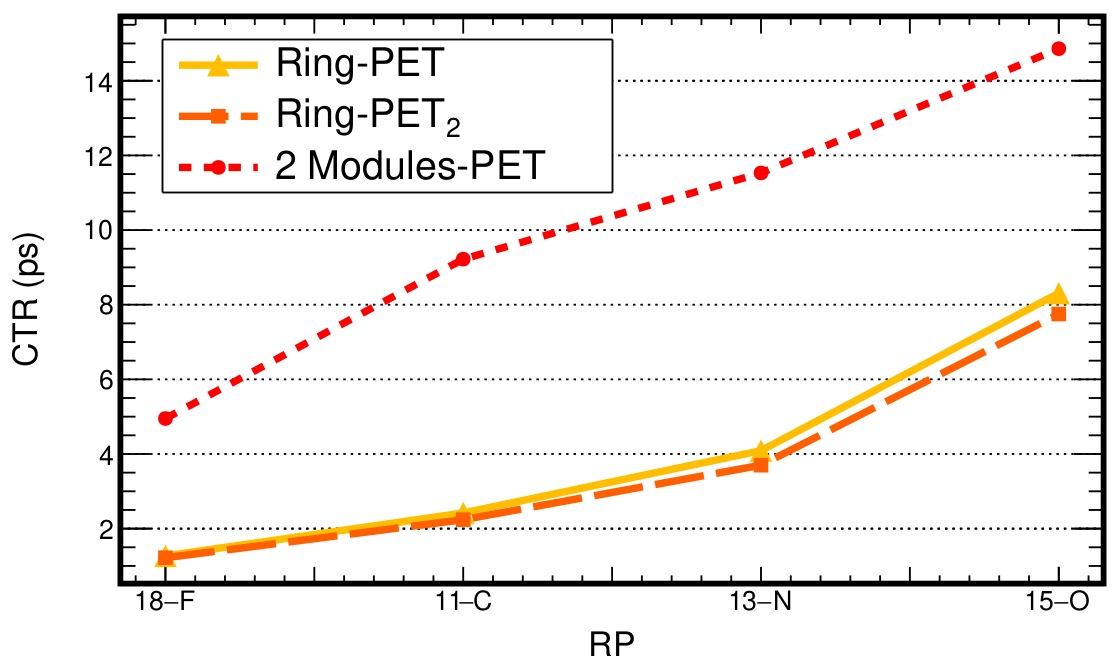}
\caption{CTR value calculated for the four RNs and three configurations.} 
\label{ff5.4}
\end{figure}

The values for the four RNs and the three configurations are presented in Figure~\ref{ff5.4}. 
 
ANOVA tests showed independence of the three variables ($p=0.03$). Post hoc tests showed that there was no statistically significant difference between CTR values of the Ring-PET and Ring-PET2 ($p=0.076$) configurations. Significant differences were obtained with the 2 Modules-PET, where the CTR values were higher by $\sim 3.68 ps$ for $^{18}$F and $\sim 7.11$ ps for $^{15}$O compared to the other two PET configurations ($p= 0.005$ for Ring-PET and $p=0.006$ for Ring-PET$_2$). These results can be seen in Table III. These calculations show that CTR values were larger for the 2 Module-PET configuration and the CTR value for the Ring-PET and the Ring-PET$_2$ was similar.

 %%%%%%%%%%%%%%%%%%%%%%%%%%%%%%%%%%%%%%%%%%
 %------------SPATIAL RESOLUTION----------%
 %%%%%%%%%%%%%%%%%%%%%%%%%%%%%%%%%%%%%%%%%%
\subsection{Spatial Resolution}
\begin{figure}[H]
\centering
\includegraphics[width=1.0\columnwidth]{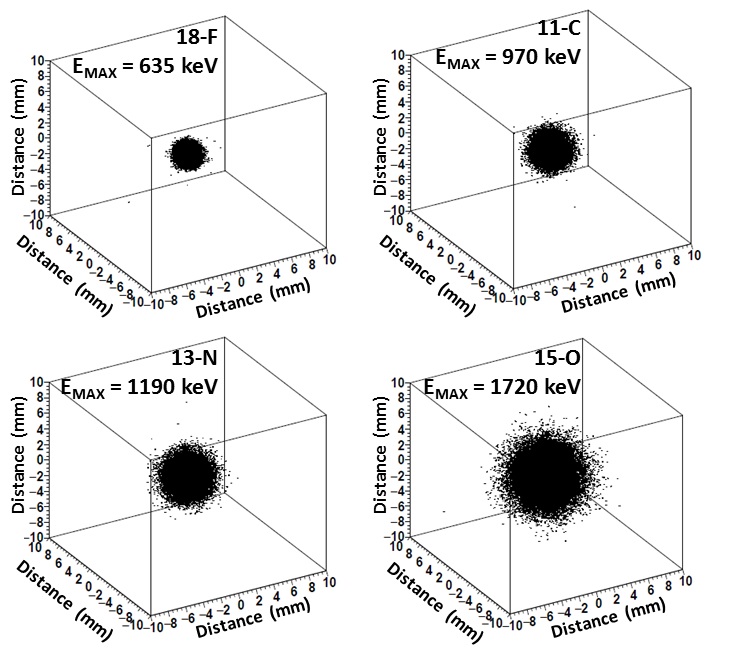}%fig7
\caption{Reconstruction of the position of annihilation events for $^{18}$F, $^{11}$C, $^{13}$N and $^{15}$O using the Ring-PET configuration.}
\label{ff5.1}
\end{figure}

Figure~\ref{ff5.1} presents a sample reconstruction of the points where the electron-positron annihilations occurred inside the water sphere with the Ring-PET configuration for the four simulated RNs.

\begin{figure}[H]
\centering
\includegraphics[width=1.0\columnwidth]{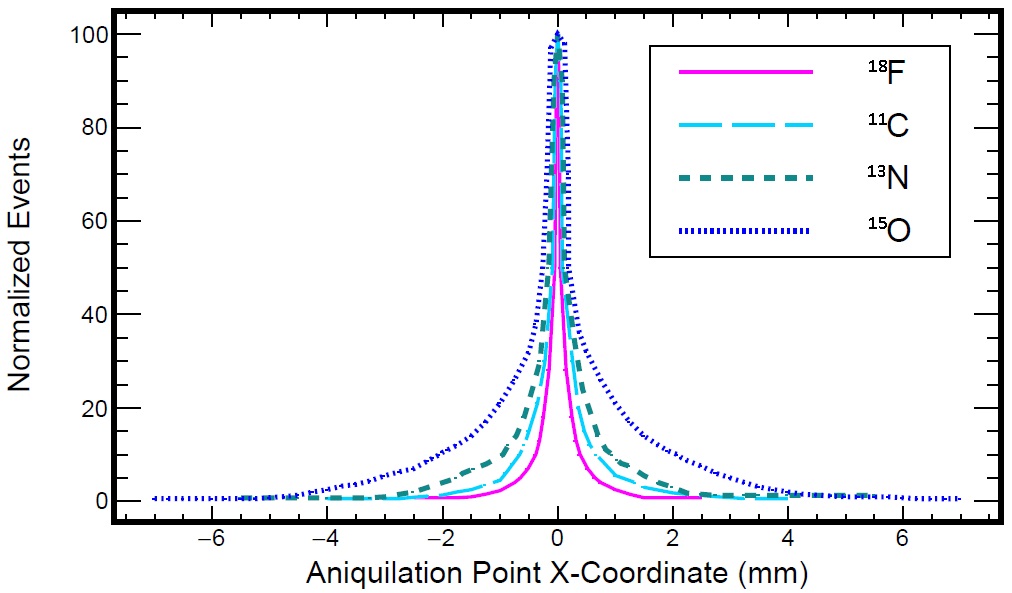}
\caption{$X$-coordinate reconstruction for Ring-PET configuration and the four RNs: $^{18}$F, $^{11}$C, $^{13}$N and $^{15}$O. Distributions were normalized for clarity.} 
\label{ff5.2}
\end{figure}

Figure~\ref{ff5.2} shows the histogram of the distribution on the $X$ coordinate for the four RNs in the Ring-PET configuration. The analysis was repeated for the three PET distributions and the four RN for comparison purposes.

\begin{figure}[h!]
\centering
\includegraphics[width=1.\columnwidth]{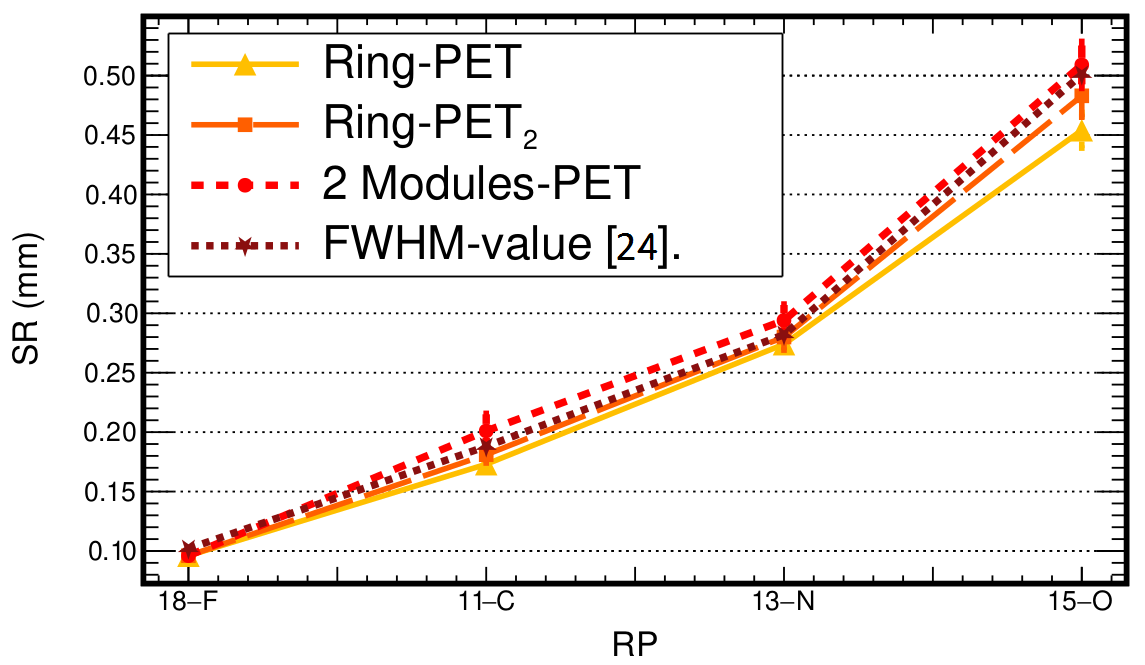}
\caption{SR calculated from the FWHM method. Values for the X-coordinate of the annihilation point for: $^{18}$F, $^{11}$C, $^{13}$N and $^{15}$O. Graph comparing the three PET configurations.} 
\label{ando}
\end{figure}

\begin{figure}[H]
\centering
\includegraphics[width=1.\columnwidth]{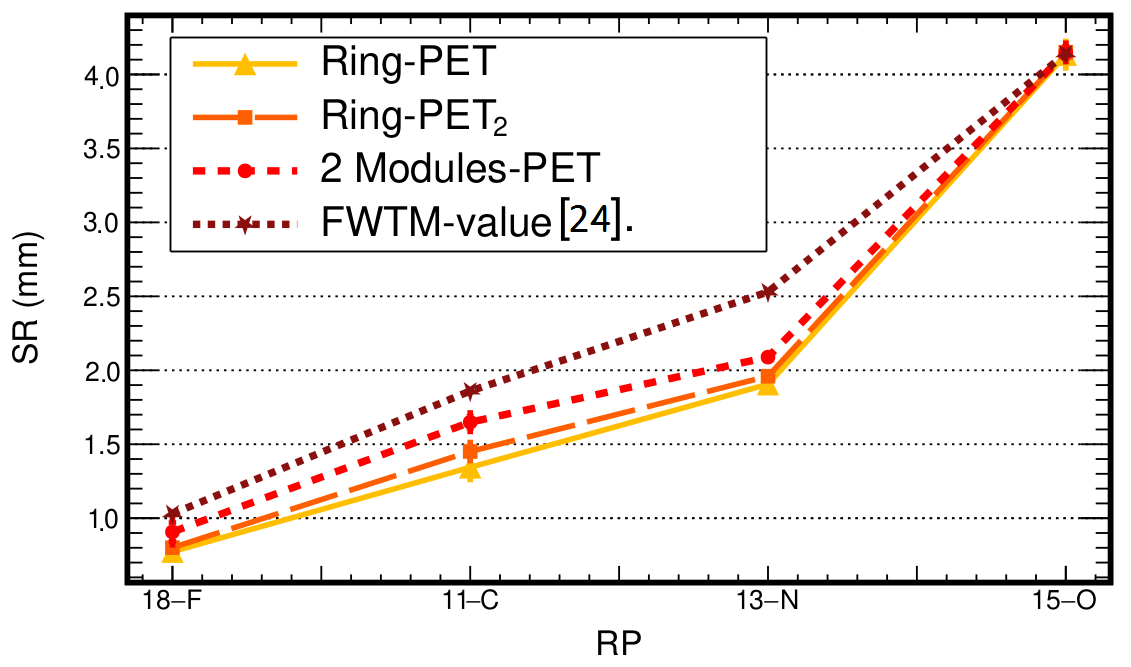}
\caption{SR calculated from the FWTM method. Values for the X-coordinate of the annihilation point for $^{18}$F, $^{11}$C, $^{13}$N and $^{15}$O. Comparison performed on the three PET configurations.} 
\label{decimo}
\end{figure}

Figures \ref{ando} and \ref{decimo} present the results from SR calculations. Both graphs are presented for the four RNs, but the first one calculated SR with the FWHM method and the second with the FWTH method.

%%%%%%%%%%%%%%%%%%%%%%%%%%
%%%%%%%%%%%%%%%%%%%%%%%%%%%%%%%%%%% FIGURA

These results can be seen quantified in Table III. For the SR comparisons, the ANOVA tests showed independence of the three variables (configuration) for the FWHM and the FWTH measurements ($p<0.001$ in both cases). The two Ring configurations showed no significant differences with each other for FWHM ($p=0.187$) and for FWTH ($p=0.089$). When comparing Ring-PET with 2 Modules-PET, FWHM differences were not significant ($p=0.087$), and FWTH were also not significant ($p=0.077$). When comparing Ring-PET2 with 2 Modules-PET, FWHM differences were also not significant ($p=0.065$), and either were FWTH differences ($p=0.078$). It is of interest to point out that almost all these values were not statistically different for small margins ($0.05<\alpha<0.1$), which shows a trend of Ring configurations having a slighter smaller (and therefore better) SR.

\subsection{Acceptance}
Table~\ref{tab5} presents the $\mathcal{A}$ values for the three PET configurations considering the four RNs inside of the modelled sphere. For the A comparisons, ANOVA tests showed independence of the three variables ($p=0.021$). Post hoc tests showed that: there was a significant difference between both Ring-PET configurations with respect to $\mathcal{A}$ values ($p=0.001$), being Ring-PET A values larger than those of Ring-PET$_2$. Ring-PET and 2Modules-PET had also significant different A values ($p=0.001$) being the Ring-PET $\mathcal{A}$ values larger. In contrast, Ring-PET$_2$ $\mathcal{A}$ values were not significantly different from those of 2 Modules-PET ($p=0.187$).

\begin{center}
\begin{table*}[htbp]
%\renewcommand{\arraystretch}{1.5}
%\resizebox{10cm}{!} {
\begin{tabular}{ | c| 
c| c| c| c|
%c| c| c| c| 
%c| c| c| c| 
%c| c| c| c| 
%c| c| c|
}
\hline
%\multicolumn{4}{ |c| }{Acceptance} \\ \hline
%RP & Ring-PET & Ring-PET$_2$ & 2 Modules-PET \\ 
RN 
& $^{18}F$ & $^{11}C$ & $^{13}N$ &  $^{15}O$
\\\hline
 & \multicolumn{4}{c|}{FWHM (mm)}
\\
 \hline
Ring-PET 
& $0.096 \pm 0.002 $ & $0.173 \pm 0.002$ &$0.274 \pm 0.009$ &$0.454 \pm 0.015$\\
Ring-PET$_2$ 
& $0.096 \pm 0.002$ & $0.181 \pm 0.007$ &$0.280 \pm 0.011$ &$0.483 \pm 0.018$\\
2 Modules-PET 
& $0.096 \pm 0.002$ & $0.201 \pm 0.014$ &$0.294 \pm 0.013$ &$0.509 \pm 0.090$\\
Literature [24] 
& $0.102$ & $0.188$ &$0.282$ &$0.501$\\

\hline
& \multicolumn{4}{c|}{FWTM (mm)}
\\
 \hline
Ring-PET 
& $0.774 \pm 0.037$ & $1.344 \pm 0.082$ &$1.906 \pm 0.058$ &$4.134 \pm 0.090$
\\
Ring-PET$_2$ 
& $0.801 \pm 0.040$ & $1.451 \pm 0.061$ &$1.960 \pm 0.033$ &$4.149 \pm 0.056$
\\
2 Modules-PET 
& $0.907 \pm 0.080$ & $1.649 \pm 0.056$ &$2.089 \pm 0.044$ &$4.149 \pm 0.056$
\\
Literature [24] 
& $1.03$ & $1.86$ &$2.53$ &$4.14$\\
\hline
 & \multicolumn{4}{c|}{CTR (ps)} 
\\\hline
Ring-PET 
& $1.27 \pm 0.06$ & $2.42 \pm 0.01$ &$4.09 \pm 0.02$ &$8.31 \pm 0.04$
\\
Ring-PET$_2$ 
& $1.22 \pm 0.01$ & $2.24 \pm 0.02$ &$3.70 \pm 0.04$ &$7.75 \pm 0.06$ 
\\
2 Modules-PET 
& $4.95 \pm 0.08$ & $9.22 \pm 0.13$ & $11.53 \pm 0.14$ & $14.86 \pm 0.17$
\\\hline
 & \multicolumn{4}{c|}{$\mathcal{A} (\%)$} 
\\
 \hline
Ring-PET 
& $17.41 \pm 0.96$ & $17.15 \pm 0.97$ &$16.95 \pm 0.98$ &$16.54 \pm 1.01$
\\
Ring-PET$_2$ 
& $14.54 \pm 1.15$ & $14.50 \pm 1.15$ &$14.50 \pm 1.15$ &$14.59 \pm 1.14$ 
\\
2 Modules-PET 
& $14.49 \pm 1.15$ & $14.12 \pm 1.18$ & $13.85 \pm 1.21$ & $12.97 \pm 1.29$
\\\hline
\end{tabular}
%}
\caption{SR, CTR and Acceptance values for the three PET configurations and the four RNs.}
\label{tab5}
\end{table*}
\end{center}

 %%%%%%%%%%%%%%%%%%%%%%%%%%
 %%%%%%%%%%%%%%%%%%%%%%%%%%%%%%%%%%%%%%
\section{Discussion}
In this work we modelled using Geant4, three configurations for possible preclinical PET scanners with pairs of LYSO + SiPM of different sizes: This to find which configuration would be best in future scanners. The SR values calculated, were in the range of previously reported SR values in the literature (24), but did not present differences between the three configurations. CTR values were statistically better for the Ring-PET configurations compared to the 2 Modules-PET while CTR values for Ring-PET and Ring-PET$_2$ were similar. The Ring-PET configuration due to its larger detection area, was more efficient in capturing gammas.

\subsection{CTR}
As mentioned in previous sections, positrons with higher kinetic energy (1,720 keV) reach greater distances before the annihilation process, compared to positrons with lower kinetic energy (635 keV) \cite{carter}. Therefore, the annihilation point was located at a greater distance from the center of the sphere, closer to one crystal than the diametrically opposite (see Figure \ref{ff5.1}). Thus, the difference in the arrival time of the gammas to the crystals is greater for sources that produce positrons with higher kinetic energy. This behavior was observed for the three different PET configurations and for all RNs. The smallest CTR values were obtained for Ring-PET$_2$, followed by Ring-PET (see Table \ref{tab5}). The highest CTRs were found for the 2 Modules-PET, which made it, regarding the CTR parameter, the lest adequate configuration of the three for preclinical PETs.

\subsection{Spatial Resolution}
In Figure \ref{ff4.3} the maximum kinetic energy value for each RN can be observed. This characteristic turned out to be important for the annihilation point reconstruction. Positrons emitted by the $^{15}$O (EMax = 1,720 keV) traveled at least twice the distance before the annihilation process, compared with positrons emitted by $^{18}$F (EMax = 635 keV), in which the annihilation process occurred before. Consequently, the range of the positron increased as the maximum energy of the radionuclide increased. This was verified in Figure \ref{ff5.2}, in which the width of the distributions increased with energy emitted by RNs.

As it can be observed in Table \ref{tab5}, the values reported in literature (no error information available). Differences between literature values and FWHM were in average $8.9\%$, while differences for the FWTH were much larger representing $26\%$. We did not find a statistical difference between these values for the FWHM ($p=0.867$), and the same for the FWTH ($p=0.765$). Figures \ref{ando} and \ref{decimo} implement a trace that corresponded to the literature values, presented there to compare with those of the simulations presented in this manuscript. This lack of statistical difference between the results from this work and those of literature, represents a partial validation for the use of Geant4 for these kinds of modellations.

The values obtained for the Ring-PET were different to the other two PET configurations regarding FWHM. That was $\sim0.026$ mm for the Ring-PET2 and $\sim0.055$ mm for the 2 Modules-PET.

Meanwhile, the difference between the FWTM values was $\sim0.305$ mm for Ring-PET$_2$ and $\sim0.198$ mm for 2 Modules-PET. Furthermore, there was no significant evidence that the variances were different between them. Therefore, the spatial resolution was not a valid parameter in these simulations to make a decision on which configuration was best. Still, and as commented in the results section, there was an almost significant trend for the 2 Modules configuration to have a worst SR.

\subsection{Acceptance}
A values are presented in Table \ref{tab5} for the three PET configurations and the four RNs. The 2 Modules-PET had a smaller detection area: therefore, it had the lower $\mathcal{A}$ value, being  $\sim$14.49 and  $\sim$12.97 for $^{18}$F and $^{15}$O respectively. The largest $\mathcal{A}$ value was obtained for the Ring-PET, (with $\sim$17.41 and $\sim$16.54 for $^{18}$F and $^{15}$O respectively), which had bigger crystal sizes than the Ring-PET$_2$. This made it a more feasible configuration to detect gamma particles. It can be concluded that this parameter depends mainly on detector area and for these configurations Ring-PET was the best as it had larger LYSO detecting elements.

\subsection{Limitations}
The CTR values presented in this work are at most 15 ps for the 2 Modules-PET and the $^{15}$O RN. One the of the limitations of this study was to be able to physicaly measure events in that time order. A typical Time to Digital Counter (TDC) has a time resolution of 20 ps. Typically CTR values are in ns \cite{capilar}, which were i.e. the 2 Modules-PET values. However, a new pico-TDC card has recently been designed to measure 2.2 ps \cite{B28}. Then, it will be possible to measure the values obtained in this work.

%%%%%%%%%%%%%%%%%%%
%%%%%%%%%%%%%%%%%%%%%%%%%%%%%%%%%%%%%%%
\section{Conclusions}\label{conclusions}
n this work we modelled using Geant4, three configurations for possible micro-PET scanners with pairs of LYSO + SiPM of different sizes: all this for four RNs of typical use in nuclear medicine. The main objective of this work was to facilitate future PET designs and construction saving time and resources. The SR values modeled here were in the range of the reported SR values in the literature \cite{bib9}. This represents a partial validation of the use of Geant4 as a platform on which these types of simulations were possible. CTR values were statistically better for the PET ring configurations compared to the 2 Modules-PET while CTR values for Ring-PET and Ring-PET$_2$ were similar. The radionuclides that produced positrons with high energy presented a higher positron range than radionuclides that emitted positrons with lower energy. Therefore, the energy of the simulated source is a factor that affects the SR and CTR parameters. The Ring-PET configuration due to its larger detection area (Acceptance parameter), was more efficient in capturing gammas, and therefore reconstructing a greater number of annihilation points to produce a better image. The result of this modelling effort and this parameter analysis was, that the 2 Modules-PET configuration was worst in its performance than the Ring configurations. When comparing Ring PET and Ring-PET$_2$ comparisons, there was a trend for the first being better (larger A but similar SR and CTR).

\section{Declaration of competing interest}
The authors declare that they have no known competing finan-
cial interests or personal relationships that could have appeared to influence the work reported in this paper.

\section{Acknowledgements}
Authors would like to thank the Consejo Nacional de Humanidades Ciencias y Tecnología (CONAHCyT) for the doctoral grant of MLLT. Authors would also like to thank the access to the facilities of the Medical Physics Lab from the Faculty of Mathematical and Physical Sciences, as well as the National Supercomputing Laboratory of Southeast Mexico (LNS) of the Benemérita Universidad Autónoma de Puebla (BUAP).

\end{document}